\documentclass[tbtags,aps,prd,twocolumn,groupedaddress,showpacs,preprintnumbers,amsmath,amssymb,floatfix]{revtex4}
\usepackage[english]{babel}

\usepackage[T1]{fontenc}
\usepackage{inputenc}
\usepackage[final]{graphicx}% Include figure files
\usepackage{subfig}
\usepackage{enumerate}
\usepackage{amsmath}
\usepackage{amsfonts}
\usepackage{amsthm}
\usepackage{amssymb}
\usepackage{xcolor}
\usepackage{calc}
\usepackage{ifthen}

\newcommand{\reals}{\mathbb{R}}
\newcommand{\sphere}{\mathsf{S}}
\newcommand{\torus}{\mathsf{T}}
\DeclareMathOperator{\union}{\cup}
\DeclareMathOperator{\homeo}{\simeq}
\newcommand{\density}[1]{\mathcal{#1}}
\newcommand{\half}{\frac{1}{2}}
\newcommand{\doo}{\partial}
\newcommand{\doomu}{\doo_\mu}
\newcommand{\dooMU}{\doo^\mu}
\DeclareMathOperator{\defin}{\equiv}
\newcommand{\norm}[1]{\lVert#1\rVert}
\DeclareMathOperator{\grad}{\nabla}
\DeclareMathOperator{\cartprod}{\times}

\newcommand{\integers}{\mathbb{Z}}
\newcommand{\naturals}{\mathbb{N}}

\begin{document}

\title{Unwinding in Hopfion vortex bunches}

\author{Juha J\"aykk\"a}
\email{juolja@utu.fi}
\author{Jarmo Hietarinta}
\email{hietarin@utu.fi}
\affiliation{
Department of Physics and Astronomy,
University of Turku,
FI-20014 Turku, Finland
}

\date{\today}

\begin{abstract}
  We investigate the behaviour of parallel Faddeev-Hopf vortices under
  energy minimization in a system with physically relevant, but unusual
  boundary conditions.  The homotopy classification is no longer provided
  by the Hopf invariant, but rather by the set of integer homotopy
  invariants proposed by Pontrjagin.  The nature of these invariants
  depends on the boundary conditions. A set of tightly wound parallel
  vortices of the usual Hopfion structure is observed to form a bunch of intertwined
  vortices or unwind completely, depending on the boundary conditions.
\end{abstract}

% ward paper has 11.27.+d, 05.45.Yv, 11.10.Lm
\pacs{11.27.+d, 05.45.Yv, 11.10.Lm}% PACS, the Physics and Astronomy
\maketitle

\section{Introduction}
The standard model with knot solitons, the Faddeev-Skyrme (FS) model, was
proposed by L.~Faddeev in 1975 \cite{Faddeev:1975tz}. Since then, many
analytical and numerical results have been obtained about this model
\cite{Vakulenko_Kapitanskii:1979, Kundu:1982bc, Faddeev:1997zj,
  Gladikowski:1997mb, Hietarinta:1998kt, Hietarinta:2000ci, Battye:1998pe,
  Battye:1998zn, Sutcliffe:2007ui}.  The localized solutions of the FS
model are characterized by the Hopf charge, which can be defined if the
field is constant at infinity: $\lim_{r\to \infty}\vec{n}(r) =
\vec{n}_\infty$; this allows one point compactification of the domain
$\reals^3 \union \{\infty\} \homeo \sphere^3$ and the definition of Hopf
charge using the homotopy group $\pi_3(\sphere^2)$. However, this is not the only
possible boundary condition.  Indeed, some physical systems, like rotating
superfluid $^3$He in the $A$ phase \cite{1995JETPL..61...49M, Ruutu:1995aa}
and certain insulating magnetic materials, called topological insulators
\cite{Moore:2008aa}, can have different boundary conditions requiring a new
topological analysis. There have already been experimental observations of
a two dimensional topological insulator in Bi$_{1-x}$Sb$_x$, having the
topological charge characterized by of $\integers_2 \times \integers_2$
\cite{D.Hsieh02132009}.

In this paper we use large scale numerical optimisation routines to
investigate the minimum energy configurations of the FS model in novel
topological situations. The topological invariants that are relevant when
the boundary conditions are no longer the usual $\lim_{r\to
  \infty}\vec{n}(r) = \vec{n}_\infty$ are discussed in section
\ref{sec:topology}.  It turns out that new topological invariants,
introduced by Pontrjagin \cite{Pontrjagin:1941}, are needed whenever some
periodic boundary conditions are used.  In section
\ref{sec:the_faddeev_skyrme_model} we describe the Faddeev-Skyrme model and
the initial configurations. In Section \ref{sec:results} we discuss the
process leading to minimum energy configurations from initial
configurations consisting of tightly wound and packed vortices parallel to
the $z$-direction.  We will demonstrate how the topological invariants of
section \ref{sec:topology} are conserved while allowing the unwinding of
the Hopfion vortices. The behaviour turns out to depend not only on the boundary
conditions, but also on the size of the periodic cell as compared to the region
occupied by the vortices.  Intermediate states of the energy minimization
process are used to illustrate the processes that unwind the knot. Locally,
these processes are quite similar to those seen in the context of closed
Hopfions \cite{Hietarinta:2002xxx}.

\section{\label{sec:topology}Topology}
We study the FS model in situations where the physical space cannot be
compactified to $\sphere^3$ but is periodic in some direction(s). In
particular we consider the case when the domain can be identified as
$\sphere^2 \cartprod \torus^1$ (periodic in the $z$-direction) or
$\torus^3$ (periodic in all directions). In these cases the field $\vec{n}$
becomes a map $\vec{n}:\sphere^2 \cartprod \torus^1 \to \sphere^2$ or
$\vec{n}:\torus^3 \to \sphere^2$, respectively.  Here we differentiate
between $\sphere^1$ and $\torus^1$: If there is a special point $x^*$ such
that $f(x^*)=f(-x^*)$ (e.g. the point at infinity) we use $\sphere^1$ while
if the periodicity can be expressed by $f(x)=f(x+L),\,\forall x$ then we
use $\torus^1$.

Note that computationally, the difference between the three cases above is
only that of the boundaries. If we want to model the $\sphere^3$ boundary
condition $\lim_{r\to \infty}\vec{n}(r) = \vec{n}_\infty$ then in the
computational lattice the fields at the edges of the box will be fixed to
$\vec{n}_\infty$ and no interaction can occur across the edges.  Next
suppose we use the same initial confinguration, but during
minimization we only enforce the periodicity condition $\vec{n}(x_i^{max})
= \vec{n}(x_i^{min})$ for some $i$. If the behavior of the system is such
that the values of the field on the edges of the box do not change much the
system will behave essentially as it behaves when the $\sphere^3$ boundary
condition is used.  As we shall see, this is also reflected by the fact
that in the $\torus^3$-case, the homotopy classification of
the system is still given by the Hopf invariant even though the domain is not
topologically $\sphere^3$.  The above systems behave differently only when
there is strong interaction across one or more edges of the periodic box. 

The homotopy classification of the above cases is more complicated than the
usual case of $\sphere^3 \to \sphere^2$ and was first proposed by
Pontrjagin \cite{Pontrjagin:1941}. A more modern approach, restricted to
closed, connected, oriented 3-manifolds is due to Auckly and Kapitanski
\cite{Auckly:2005aa}. We will further restrict their result to maps from
$\sphere^2 \times \torus^1$ and $\torus^3$ to $\sphere^2$. The following
geometrical descriptions are due to Kapitanski
\cite{Kapitanski_Durham:2005}.

Let $\vec{n}:\sphere^2 \times \torus^1 \to \sphere^2$. The topological
properties of $\vec{n}$ follow the topological properties of its
restrictions $\vec{n}|\sphere^2$ and $\vec{n}|\torus^1$. If we denote these
restrictions by $f$ and $g$, respectively, we have $f: \sphere^2 \to
\sphere^2$ and $g:\torus^1 \to \sphere^2$. The topological properties of
$f$ are described by the usual homotopy classification of maps $\sphere^2
\to \sphere^2$, i.e., by $\pi_2(\sphere^2)$ and therefore the relevant homotopy
invariant is $s_2 \defin \deg f \in \integers$. The topology of $g$ is more
subtle. It turns out that for each value of $s_2$ there is another
invariant, here called $s_1$, which is defined by the map $g$. It can be
shown \cite{Kapitanski_Durham:2005} that $s_1 \in \integers_{2s_2}$. 

As an example, consider the stereographic coordinate $z$ of $\sphere^2$ and
the map $\vec{n}:\sphere^2 \times \torus^1 \to \sphere^2$ for which $f(z) =
z^m$, where $m \in \naturals$. It can be shown \cite{Manton:2004tk} that
for such a map, $\deg f = m$. Thus $s_2$ is the number (with
multiplicities) of zeros of $f$, i.e. the number of vortex cores.  Now, for
a simple vortex $s_2 = \deg f = 1$ and therefore $s_1 \in \integers_2$.
Thus, there are two homotopy classes, represented by $z e^{i s_1 \theta}$,
where $\theta \in [0,2\pi]$ is the coordinate of $\torus^1$. The function
$g$ now describes how many full turns the $\sphere^2$ is rotated as
$\theta$ goes from $0$ to $2\pi$. The domain can be regarded as a set of
co-centric $2$-spheres for different values of $\theta$ with the inner- and
outermost spheres identified. (If $s_1 = 0$, $g$ is the constant map
$\theta \mapsto 0$ and $f(w,\theta)=w, \, \forall \theta$ and there is no
rotation.) The full map $\vec{n}:\sphere^2 \times \torus^1 \to \sphere^2$
represents an $1$-vortex, where two distinct preimages of points on the
target $\sphere^2$ have linking number $H$, that we still call the Hopf
invariant, but it is not a homotopy invariant in this case. 
% \juhaj{Consider the point $p=(1,0) \in \sphere^2$. The stereographic
%   projection takes this to $w=1$ and also $f(w,0)=1$.  If $s_1 = 1$, the
%   map $\vec{n}$ moves this point around $\sphere^2$ when $\theta$ goes
%   around the $\torus^1$: $f(w,\pi/2) = i$, $f(w,\pi)=-1$,
%   $f(w,2\pi/3)=-i$ and finally $f(w,2\pi)=f(w,0)=1$.

Detailed description of how the higher numbers of rotations can in this
case be unwound to either $s_1 = \Bar{0}$ or $s_1 = \Bar{1}$ can be found
in \cite{Kapitanski_Durham:2005}, but we will not discuss this further.
The case $s_2=0$ is rather special and is not needed here.

Next, consider $\vec{n}:\torus^3 \to \sphere^2$. According to Pontrjagin
\cite{Pontrjagin:1941}, there are three primary homotopy invariants, $t_1$,
$t_2$ and $t_3$. In order to determine the values of $t_i$, consider
$\vec{n}^{-1}(p) \in \torus^3$, the preimage of a \emph{regular value}
(under the map $\vec{n}$) $p\in\sphere^2$. By continuity of $\vec{n}$, the
preimage is a closed loop and therefore can be represented by a directed
path $\gamma:\sphere^2 \to \torus^3$.
%, with the direction defined by the sign of $\tfrac{\partial \gamma(u)}{\partial u}$.
Now, $\gamma \in \pi_1(\torus^3) = \integers \cartprod \integers \cartprod
\integers$ and each of the invariants $t_i$ belong to one of the
$\integers$ of $\pi_1(\torus^3)$. Visually, $t_i$ count the number of times
$\gamma$ travels around the respective $\torus^1$ of the $\torus^3$ and can be
determined as follows:
For each $t_i$, find all the points (with multiplicities) where $\gamma$
pierces a plane defined by $\Hat{e}_i$, numbering them with $j \in
\integers$. For each of these points, let $\epsilon(j) = +1$ if $\gamma$ is
directed in the same direction as $\Hat{e}_i$ and $\epsilon(j) = -1$ if the
directions are opposite. (Note that by ``piercing'' we mean that the
preimage has to go through the plane, not just touch it at a point.)  Now,
for each $i$, we define $t_i := \sum_j \epsilon(j)$.
%
%\juhaj{If we keep the third coordinate on the $\torus^3$ fixed and consider
%  just the map $\vec{n}\suchthat \torus^2$, the resulting integers $t_1$
%  and $t_2$ are exactly the integers describing the torus knot
%  $\mathcal{K}_{t_1t_2}$. Restriction to the other pairs of coordinates
%  produce torus knots similarly. \textbf{((** Vielä lisäkommentti
%  torussolmuista: Jos $1<T=\gcd{t_1,t_2}$, torussolmut
%  $\mathcal{K}_{t_1t_2}$ ja $\mathcal{K}_{t_1/T,t_2/T}$ ovat
%  samanlaiset. Onkohan tällä siis jotain yhteistä alla määriteltävän $t$:n
%  kanssa? Nythän $t$ jossakin mielessä identifioi ylläkuvatun kaltaisesti
%  samanlaiset 3-torus-solmut antamalla niille saman sekundäärisen
%  invariantin: $t=\gcd{2,4,10}=\gcd{4,8,20}=2$. Näin syvälle meneminen ei
%  taida kuulua kuitenkaan tähän**))}
%}

In the $\vec{n}:\torus^3 \to \sphere^2$ case there is also a secondary
homotopy invariant, which further divides the maps with equal $t_1$, $t_2$,
$t_3$. For fixed values of $t_i$, one defines $t = \gcd(t_1,t_2,t_3)$ and
for each $t$ there is a secondary homotopy invariant, denoted $h_t$. If $t
\neq 0$, it can be shown that $h_t \in \integers_{2t}$ \cite{Auckly:2005aa,
  Kapitanski_Durham:2005}, and thus for fixed $t_i$ and $t>0$ there are
$2t$ different homotopy classes.  If $t=0$, this invariant is exactly the
Hopf invariant and we denote it by $h_0$.

Two homotopic maps $\sphere^2\cartprod \torus^1 \to \sphere^2$ have the
same invariants $s_i$ and two homotopic maps $\torus^3 \to \sphere^2$ have
the same invariants $t_i$ and $h_t$, but due to the secondary invariants
$s_1$ and $h_t$, maps with the same $s_2$ or $t_i$ are not
necessarily homotopic. In this work, we only deal with the primary
invariants $t_i$ and $s_2$.

%For technical reasons, the core is represented by the isosurface
%of defined by the preimage of $n_3=-1+\epsilon$, where $\epsilon$ is very
%small. The preimage of $n_3=0$ is also described by a small tube is plotted
%around the curve.  We have chosen the directions of the preimages so that
%for a single vortex of the initial configuration, $t_3=1$ -- it is again
%easy to see this from figure \ref{fig:single_initial}, where the preimage
%pierces the $xy$-plane exactly once.
%
%Note that in the numerical work reported here, in two of the initial
%configurations the preimage of $(0,0,-1)$ is a set of four disconnected
%paths. These are constructed from four identical single vortices and
%therefore their directions are identical.  The directions must be assigned
%consistently when the number of disconnected paths changes during the
%minimisation, since the signs of the invariants $t_i$ are determined by
%direction.

% DEGREE
%\newcommand{\jmat}[1]{\vec{#1}}
%\begin{align}
%  \textrm{deg}(\jmat{\vec{n}}) = \frac{1}{8 \pi} \int_{\mathbf{R}^2} 
%  \texttt{d}x^2\, \epsilon_{ij}\, \jmat{\vec{n}} \cdot (\partial_i \jmat{\vec{n}} \times 
%  \partial_j \jmat{\vec{n}}) \textrm{,}
%\end{align}

%%% ISOMORPHISM: H_k(M;\integers) = H^{\dim M-k}(M;\integers), Hatcher p. 231.
%%% OTOH: H_0(T^3,\integers) = H_3(T^3,\integers) = \integers and
%%%       H_1(T^3,\integers) = H_2(T^3,\integers) = \integers^3

\section{\label{sec:the_faddeev_skyrme_model}The Faddeev-Skyrme model}
The explicit form of the Lagrangian of the Faddeev-Skyrme model can be
written in terms of a unit $3-$vector $\vec{n}$ as
\begin{align}
  \label{eq:fs_model}
  \density{L} &= c_2\doomu \vec{n}^T \dooMU \vec{n}
  + c_4 F_{\mu\nu} F^{\mu\nu},\\
  F_{\mu\nu}
% &\defin \half \vec{n}^T \doomu \vec{n} 
%\crossprod \partial_\nu \vec{n} \quad \text{or, alternatively}\\
  &= \half \epsilon_{abc} n^a \doomu n^b \partial_\nu n^c,
\end{align}
where $c_2$ and $c_4$ are coupling constants. Choosing the usual metric
$(+,-,-,-)$ yields the static energy density
\begin{align}
  \label{eq:faddeev_skyrme_energy}
  \density{E}_{FS} &\defin c_2 \norm{\grad \vec{n}}^2 + c_4 \norm{F_{jk}}^2.
\end{align}
If one requires the field to have a finite energy in $\reals^3$, it is
necessary to impose the boundary condition
\begin{gather}
  \label{eq:usual_boundary}
  \lim_{r\to \infty}\vec{n}(r) = \vec{n}_\infty = \text{ constant.}
\end{gather}
However, if is the physical space is periodic in one or more dimensions, we
have in the periodic direction(s) $x_i$ $\vec{n}(x_i) = \vec{n}(x_i+L)$, where
$L$ is the length of the periodicity.  The space no longer compactifies to
$\sphere^3$, but depending on the number of periodic dimensions, it becomes
i) $\sphere^2 \cartprod \torus^1$ if one direction is periodic and the two
others have the usual boundary condition Eq.~\eqref{eq:usual_boundary}, ii)
$\torus^2 \cartprod \sphere^1$, in the case where two directions are
periodic, and iii) $\torus^3$ if all directions are periodic. The homotopy
classifications of such fields were introduced in section
\ref{sec:topology}. We will not consider case ii) in this work.

The initial configuration of a single vortex is the same as in
\cite{Hietarinta:2003vn}. Using cylindrical coordinates $\rho,\theta,z$ and
two integers $m,n$, the form of the field $\vec{n}$ is
\begin{align}
  \label{eq:vortex_configuration}
  \vec{n} &= 
  \begin{pmatrix}
    \sqrt{1-f(\rho)^2}\,\cos(m\theta+2\pi n z/L_z)\\
    \sqrt{1-f(\rho)^2}\,\sin(m\theta+2\pi n z/L_z)\\
    f(\rho)
  \end{pmatrix},
\end{align}
where $L_z$ is the box length in the $z$-direction and $f$ is just some
profile function with $f(0)=-1$ and $f(\rho)=+1$ when $\rho=\infty$, which
is then cut into a finite size box.

Case i) with initial states composed of a single straight tigthly wound
vortex of \eqref{eq:vortex_configuration} was studied in
\cite{Hietarinta:2003vn}. Those initial states correspond to $s_2=1$ and
values of the Hopf invariant larger than one. In principle they are
homotopic to configurations of lower Hopf invariants, namely $H \in
\{0,1\}$, but there is an energy barrier which prevents the unwinding.
%In the initial configurations of that work, \juhaj{there was
%  only a single vortex and} the preimages were connected. In this paper
%will focus on \juhaj{initial configurations created by gluing together
%  multiple such initial configurations. After gluing, there are multiple
%  cores and therefore the preimages are no longer connected sets of
%  points.}

\begin{figure}[ht]
  \centering
  \includegraphics{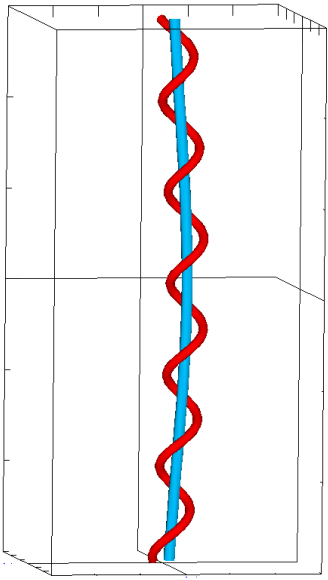}
  \caption{Single vortex in $\sphere^2 \times \torus^1$ or $\torus^3$
    domain. The vortex core $n_3=-1$ and one point from the latitude
    $n_3=0.1$ are plotted as narrow blue and red tubes, respectively. The
    wireframe indicates the boundary of the box.}
  \label{fig:single_initial}
\end{figure}

In this paper, the initial configurations are such that the preimages of
$(0,0,-1)$ are the cores of parallel vortices in the $z$-direction; one
such vortex is displayed in Figure \ref{fig:single_initial}. The small bend
visible in the figure was introduced in order to speed up the minimisation
process and does not affect the topology or linking number of the
system. Such an initial configuration has $s_2 = m$ in the case i) and
$t_1=t_2=0$, $t_3=m$ in the case iii). In both cases, the linking number of
preimages is $mn$. All of these are easy to verify visually.

Vortex bunches are constructed by putting several identical vortices into
the computational box, as shown in Figure \ref{fig:island_initial}. For the
multi-vortex configurations investigated here, we have used the same values
$n=6$ and $m=1$ for each vortex. The value of $n$ may seem rather high, but
we have learned in \cite{Hietarinta:2003vn} that for low values the the
vortex does not bend much and therefore would not properly interact with
the neighboring vortices in the current situation.

In this work, we use the same programs as in \cite{Hietarinta:1998kt,
  Hietarinta:2000ci, Hietarinta:2002xxx, Hietarinta:2003vn} because the
only difference in the computations are the boundary conditions.
% While in \cite{Hietarinta:1998kt, Hietarinta:2000ci} the boundary
% condition $\lim_{r\to \infty}\vec{n}(r) = \vec{n}_\infty$ was used, thus
% compactifying the domain to $\sphere^3$, in \cite{Hietarinta:2003vn} the
% boundary condition was $\vec{n}(x,y,z) = \vec{n}(x,y,z+L)$ and
% $\lim_{r_0^2 = x^2+y^2 \to \infty}\vec{n}(r_0) = \vec{n}_\infty$, making
% the domain $\sphere^2 \cartprod \torus^1$ and finally in this work, the
% boundaries are fully periodic, $\vec{n}(x,y,z)=\vec{n}(x+L,y,z)=
% \vec{n}(x,y+L,z)=\vec{n}(x,y,z+L).$ and the domain is therefore
% $\torus^3$.
For numerical computations, this change of boundary conditions is rather
minimal, since the MPI library directly supports both periodic and fixed of
boundary conditions defined independently for each dimension.

%\section{\label{sec:large}Vortex lattice in large domain}
\section{\label{sec:results}Results}

We now describe how an initial configuration of a bunch of Hopfion vortices
built from Eq. \eqref{eq:vortex_configuration} continuously deforms to a
minimum energy configuration while conserving the homotopy invariants
$s_i$, $t_i$. We will also describe, how, depending on the details of the
initial configuration, the linking number or the Hopfion (the Hopf
invariant) is sometimes conserved and sometimes not.

All the computations have used fully periodic boundary conditions, but in
different computational lattices.

\subsection{$2\times 2$ vortices in a large box}
Let us first consider an initial configuration consisting of $2 \times 2$
vortices parallel to the $z$ axis, put into a very large box. It is
expected that there would be negligible interaction across the edges of the
box, i.e.  the system would behave as it would in $\sphere^2 \cartprod
\torus^1$.  In the initial configuration, it was necessary to pack the
vortices into a tight formation with enough twisting ($n=6$) to ensure that
they interact, but not across the boundary.  Technically, this was
accomplished by building the initial configuration from $16$ narrow boxes:
the four central ones contained a single vortex each, while the remaining
$12$ pieces were filled with the vacuum, $\vec{n}=(0,0,1)$. In contrast to
\cite{Hietarinta:2003vn} we do not fix the field value at the $x,y$
boundary but only require periodicity, this allows us to see whether the
behavior of the vortex bunch is the same in $\sphere^2 \cartprod \torus^1$ and
$\torus^2 \cartprod \torus^1$, where the $\torus^2$ is the large box in the
$xy$-plane. Indeed, the values on the $xy$-boundaries do not change much
during the minimisation process.

\begin{figure*}[ht]
  \centering
  \subfloat[fig:island_initial][Initial state of $4$ vortices with $n=6$
  each \label{fig:island_initial}]
  {\includegraphics[]{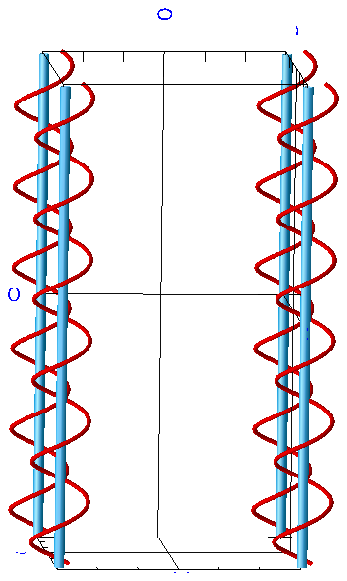}}
  \,
  \subfloat[][The preimages are about to touch at ($50000$ iterations) 
  \label{fig:island_touch}]
  {\includegraphics[]{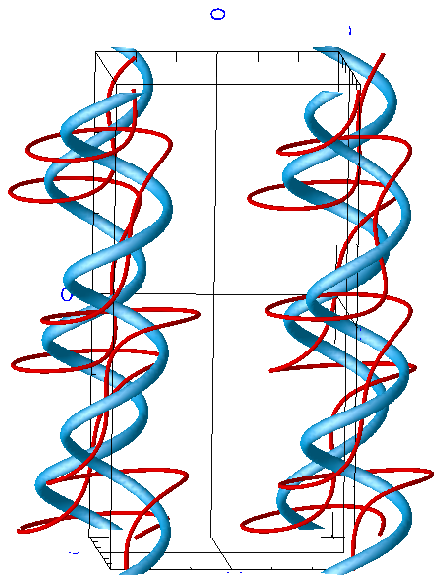}}
  \,
  % \\
  \subfloat[][The reconnected preimages have joined to vortices together
  ($60000$ iterations) \label{fig:island_reconnect2}]
  {\includegraphics[]{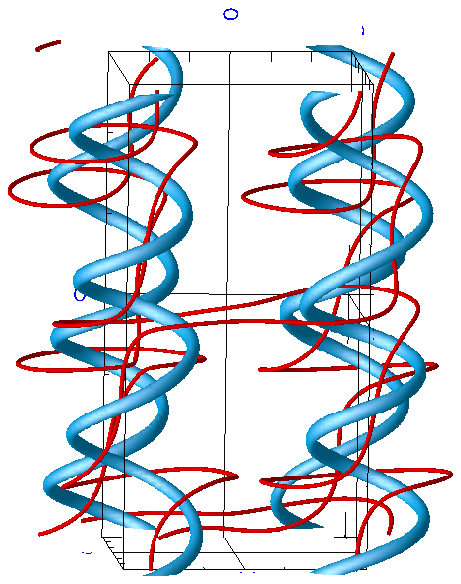}}  
  \,
  \subfloat[][Finally, (after 300000 iterations) the whole set of vortices
  is bundled into a single bunch by individual preimages wrapping around
  all the cores. \label{fig:island_bunch}]
  {\includegraphics[]{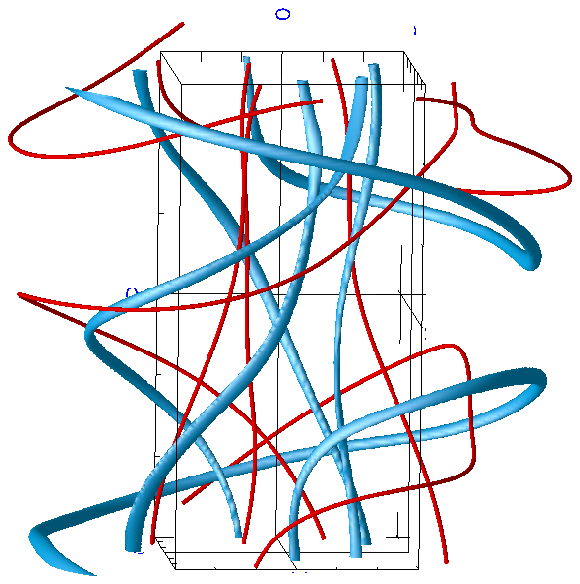}}
  \caption{Relaxation of an initial configuration of four vortices in
    $\sphere^2 \cartprod \torus^1$ domain. The colours are as in Fig.~
    \ref{fig:single_initial}, but the wireframe now shows the size of the
    lattice in $z$-direction and the exact locations of the vortex cores
    (the vertical wires). The computational box is larger and the thicknesses
    of the tubes are arbitrary.}
  \label{fig:island_all}
\end{figure*}

The initial configuration is shown in Fig.~\ref{fig:island_initial}. Since
$\vec{n}_\infty = (0,0,1)$, as seen from
Eq.~\eqref{eq:vortex_configuration}, the vortex core is at $\vec{n} =
(0,0,-1)$. It is easy to count the linking number of preimages, which
equals to $24$ (recall that there is a $\half$ factor involved -- there are
$48$ crossings) and also the homotopy invariants $t_i$: $t_1$ and $t_2$ are zero
(no winding the in $x$- and $y$-directions) and $t_3=4$ (four cores pierce
the $xy$-plane).

Energy minimisation is applied to the initial configuration using a
gradient based algorithm \cite{Hietarinta:1998kt}.  Initially, the interaction between
vortices is negligible, but after around $40000$ iterations, certain parts
of different vortices are close enough to produce evolution which differs
from that of a single vortex. At $50000$ iterations, these parts touch and
the corresponding preimages reconnect, joining two vortices together.
Fig.~\ref{fig:island_touch} shows the situation before and
\ref{fig:island_reconnect2} just after the reconnection. The wireframe box
showing the initial locations of the vortex cores is also displayed for
reference. The reconnection process is allowed because mathematically, the
red curve is a single, albeit multiply connected, preimage, and
the elementary process is exactly the same as is described in
\cite{Hietarinta:2002xxx} for a single vortex. Such deformation processes
repeat several times in different parts of the lattice until after $300000$
iterations, the initially separate vortices have formed a tangled bunch
where the cores (and preimages of a point where $n_3=0$) wind around each
other (see Fig.~\ref{fig:island_bunch}). Visual inspection of the
intermediate configurations confirms that, as expected, there is no
interaction across the $xy$-boundary. Note that there are six cores (and
preimages) piercing the the box at the top and bottom edges, but the
direction of one of them is opposite to the others, giving $t_3=4$ as
required.

The final, minimum energy configuration, has the same linking number of
preimages and integers $t_i$ as the initial state. This is despite the fact
that the linking number is no longer guaranteed to be conserved. Its
conservation is solely due to the boundaries of the box being very far from
the vortices. Indeed, in the next part, we will see that nonconservation is
possible.

\subsection{Vortices in a small box\label{ss:22s}}
In this case the same initial configuration is put into a box half the
previous size, with the expectation that there now would be interaction
across the edges of the box. Technically, this was accomplished by taking
the same four vortices as above but omitting the $12$ vacuum pieces. Again,
the $x,y$-directions are periodic but now the vortices are equally spaced
with no change in the spacing at the boundary of the computational box.

The beginning stages of the energy minimization process are identical to
that of the large box, but when the vortices start to feel each other the
configuration developes in a different manner. Instead of reconnecting to
join two vortices together, the reconnected preimages tend to develop into
straight filaments extending through the whole periodic lattice. Later,
some of the filaments again reconnect and form loops, but since these loops
have trivial linking with all other preimages, they can be deformed into
nothing.

\begin{figure*}[!Hhtb]
  \centering 
  \subfloat[][]
  {
    \includegraphics[width=4cm]{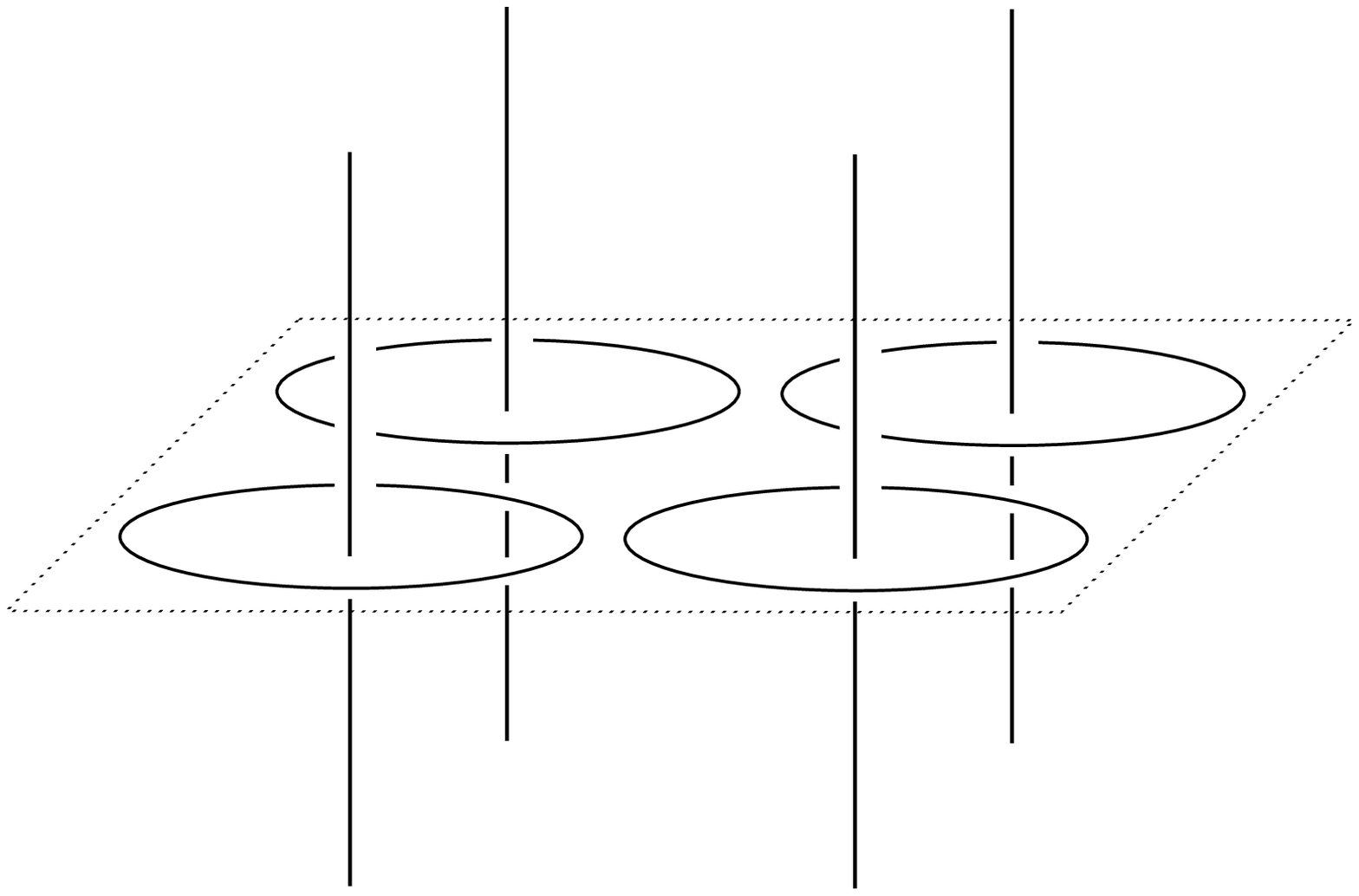}\label{fig:schematic_filament_initial}
  }
  \,
  \subfloat[][]
  {\includegraphics[width=4cm]{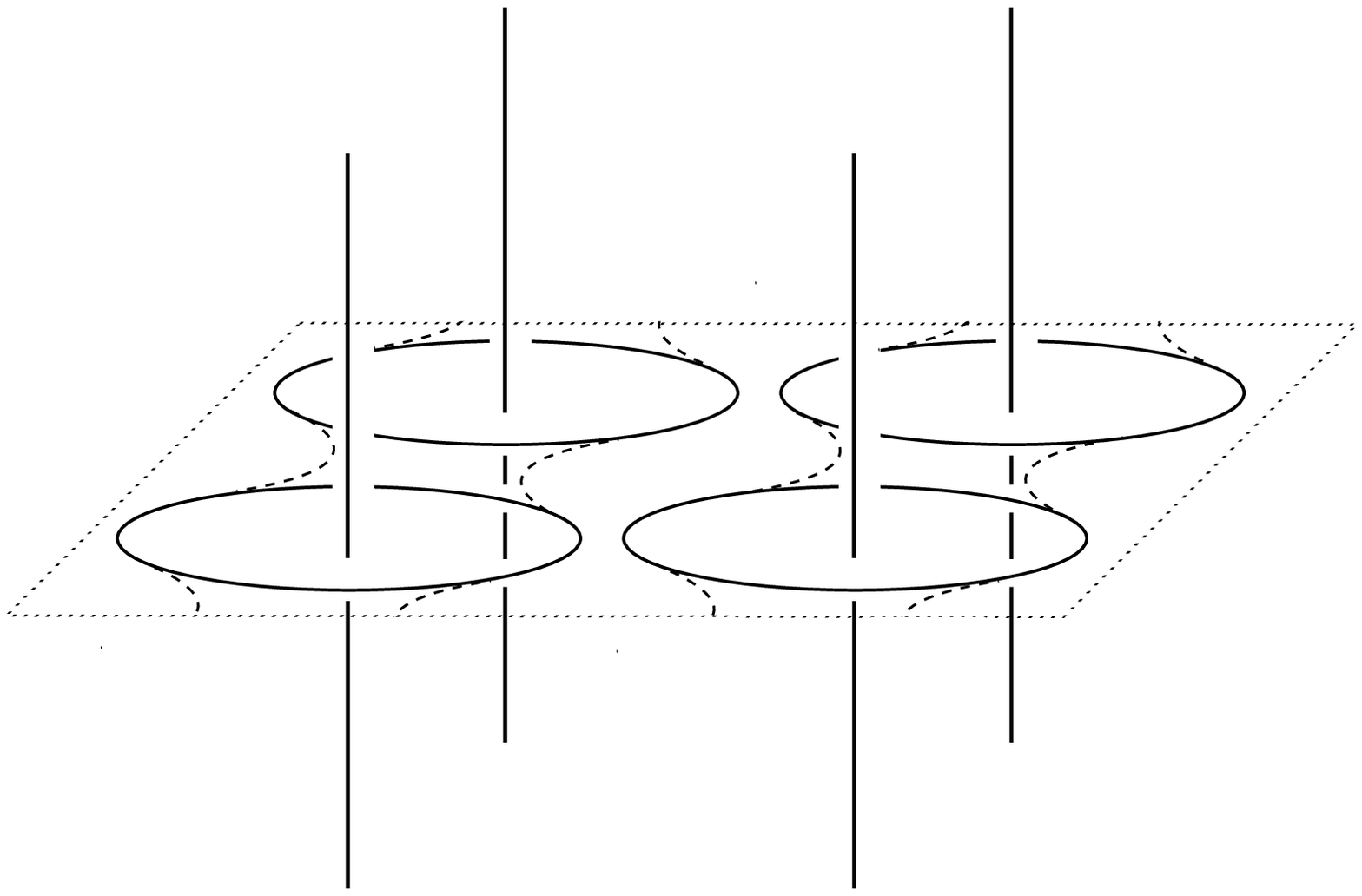}\label{fig:schematic_filament_start_split}}
  %\\
  \,
  \subfloat[][]
  {\includegraphics[width=4cm]{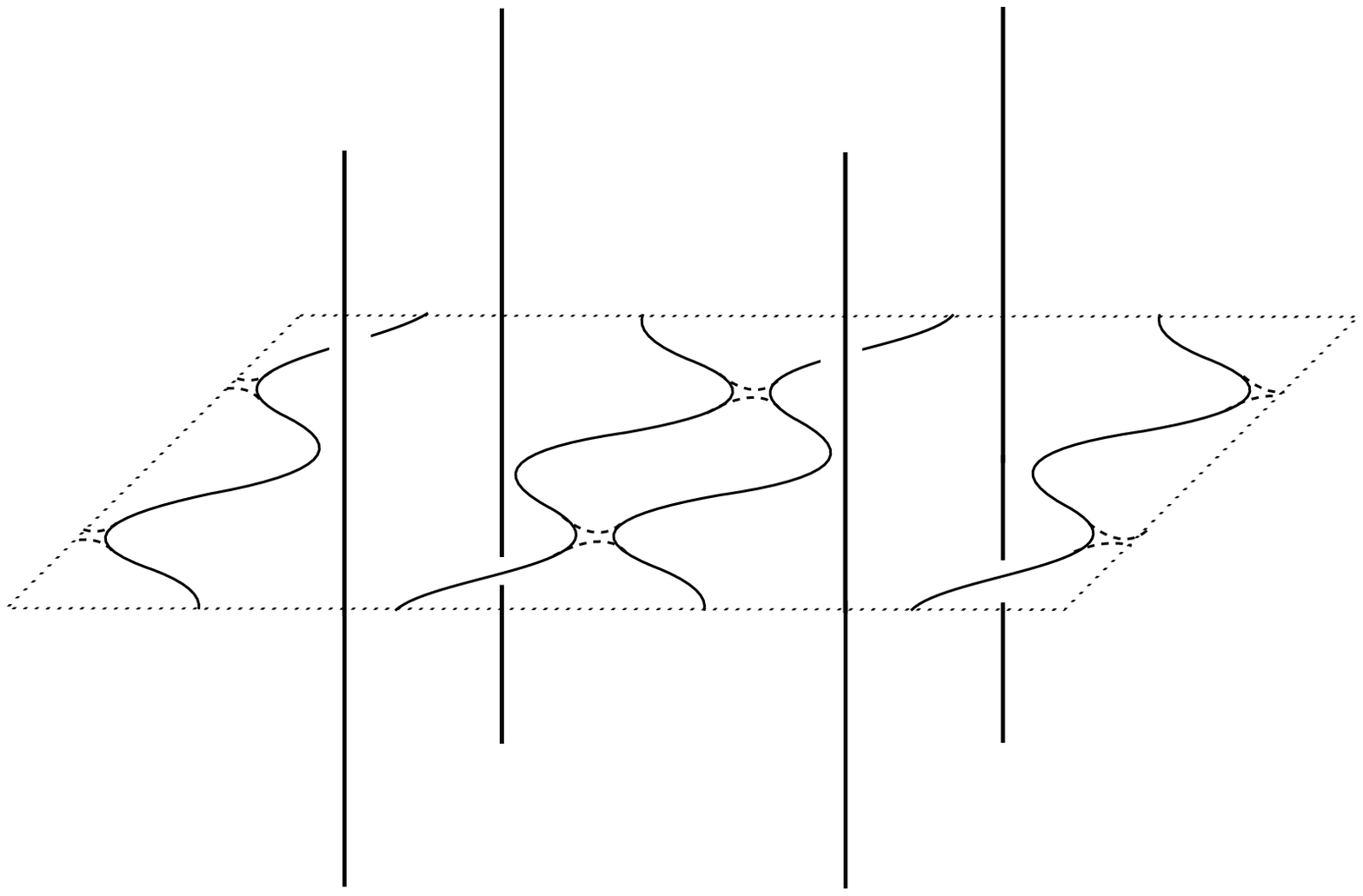}\label{fig:schematic_filament_have_split}} 
  \,
  \subfloat[][]
  {\includegraphics[width=4cm]{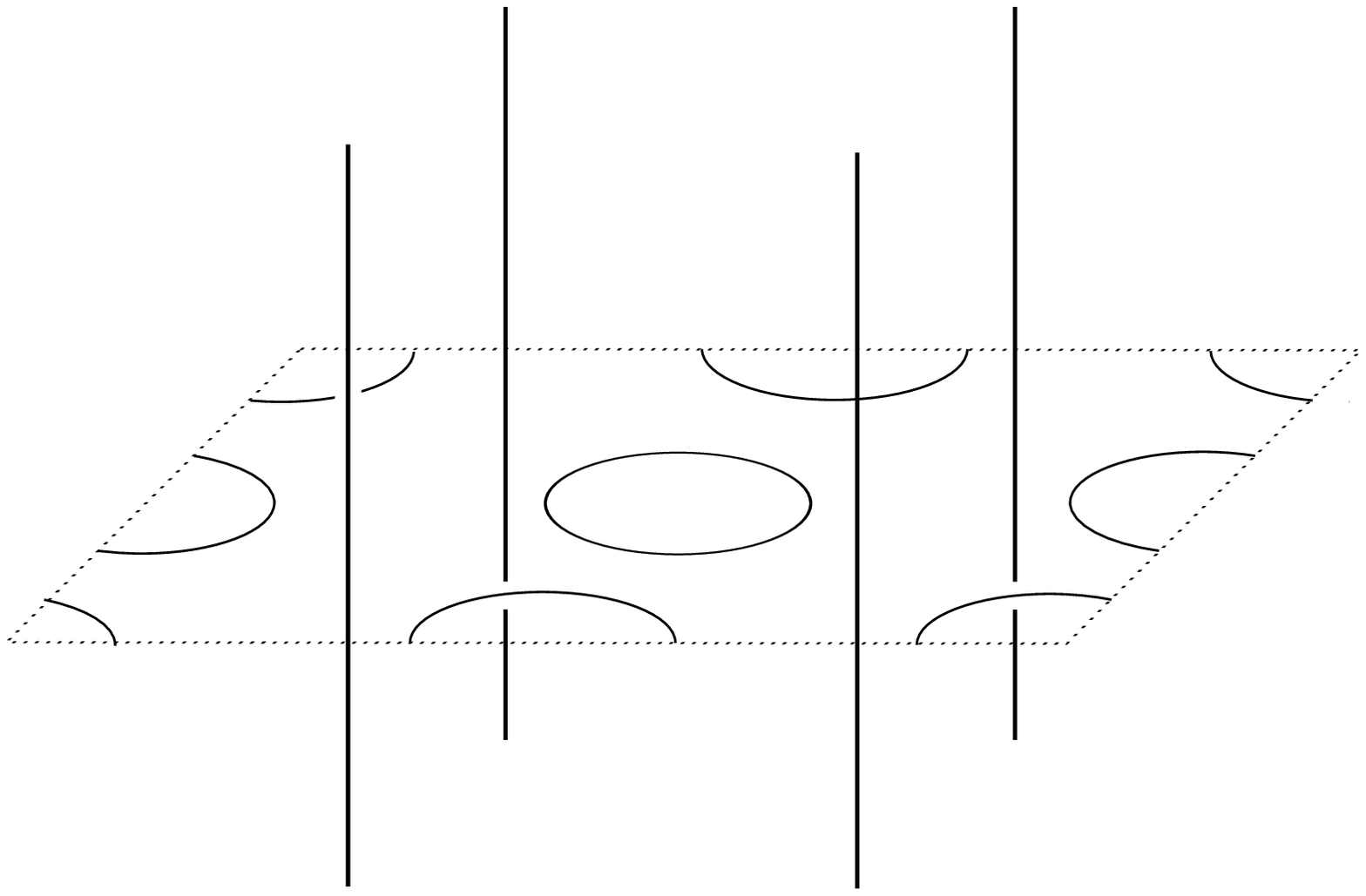}\label{fig:schematic_filament_ends}}
  \caption{Schematic illustration of the formation of filaments and
    contractible loops. The solid and dashed lines represent the preimages
    and the dotted parallelopiped is the outer boundary of the periodically
    repeating region containing the four unit cells of the initial
    vortices.}
  \label{fig:schematic_filament}
\end{figure*}

This is illustrated schematically in Fig.~\ref{fig:schematic_filament},
where initially \subref{fig:schematic_filament_initial}, processes familiar
from \cite{Hietarinta:2002xxx, Hietarinta:2003vn} have already led to each
of the four preimages to split into a relatively straight part parallel to
$z$-axis and a loop around it. The preimage loops extend, eventually
touching each other, and then reconnect along the dashed curves
\subref{fig:schematic_filament_start_split}. Upon further energy
minimisation, the filaments will in turn reconnect along the dashed curves
to form loops \subref{fig:schematic_filament_have_split}. The resulting
loops have trivial linking with other preimages and are threfore free to
collapse and vanish \subref{fig:schematic_filament_ends}. The vertical
preimages stay relatively unchanged through this process. This is the
general mechanism of Hopfion unwinding.

Indeed, the energy minimization process shows how this process of filament
and loop formation, followed by the loops shrinking and vanishing, is
repeated a number of times in different parts of the computational box so that
the final linking number of preimages is zero. The only preimages left in
the system are preimages piercing the $xy$-plane. At the same time, the
energy approaches the energy of one or more straight unwound vortices.

%A schematical illustration of the process is given in
%Fig.~\ref{fig:schematic_filament}. The
%Fig.~\ref{fig:schematic_filament_initial} displays the situation right
%before the recombination process. The curves describe a single preimage
%\juhaj{of the core, but that is not important -- they all behave similarly}
%in four unit cells of the periodic lattice. 
%
% Next, the loops are about to touch each
%other and reconnect along the dashed curves forming filaments extending
%through the whole domain. The filaments in turn will next reconnect along
%the dashed curves, forming unconnected loops in the cell centers. The loop
%is now free to shrink away. This process is repeated in different parts of
%the system eventually the Hopfion is completely unwound,
% leaving just the invariants $t_i$ unchanged.

In order to calculate the invariants $t_i$ during the minimisation process,
it is necessary to keep track of the directions of the preimages, just as
must be done for the computation of the linking number. Now, initially we
have $t_1=t_2=0$ and $t_3=4$, so whenever a preimage reaches a boundary of
the computational box and pierces it, there must always be a corresponding
piercing of the same boundary but in the opposite direction, so that $t_i$
are unchanged. Indeed, this has been observed. 
%The minimum energy (final)
%configuration is shown in Figure \ref{fig:infinite_final}, where it is easy
%to see that $t_1=t_2=0$ and $t_3=4$. Thus, $t_i$ are conserved, as
%required.

%\subsection{One vortex in a small box}
In order to further investigate the processes involved in the unwinding of
the Hopfion, we simulated a system where the
initial condition is a single vortex. Again, we use $m=1$ and $n=6$, but
instead of packing several vortices together, we take just one vortex
\eqref{eq:vortex_configuration} and add no vacuum to the system. The
boundary conditions are again fully periodic, but now the length of the
period in $xy$-directions is half of that in \ref{ss:22s}.  Any splitting
and recombination of preimages is now expected to occur across the lattice
boundary. This is indeed observed with no new processes present in the
system.

\section{Conclusions}
We have studied how the Faddeev-Skyrme model behaves when the domain
of the model is taken to be $\sphere^2 \cartprod \torus^1$ and $\torus^3$
instead of the usual $\sphere^3$. The new domains require a new homotopy
classification which is due to
\cite{Pontrjagin:1941,Auckly:2005aa,Kapitanski_Durham:2005}.  For
$\sphere^2 \cartprod \torus^1$ there is a primary homotopy invariant $s_2$
related to the $\sphere^2$ part of the field configuration and when $s_2
\neq 0$, there is also a secondary invariant $s_1$ related to the
$\torus^1$ part. We discuss the invariant $s_2$ only and it is seen to be
conserved in the energy minimisation process. For $\torus^3$, there are
three primary homotopy invariants, $t_1$, $t_2$, $t_3$ related to the
winding numbers around the three tori $\torus^1$ of the $\torus^3$. There
is a secondary invariant in this case also.  When at least one $t_i\neq 0$,
one defines $t=\gcd (t_1,t_2,t_3)$ and the secondary invariant takes values
in $Z_{2t}$. If all $t_i=0$, the secondary invariant becomes the Hopf
invariant. We have discussed the primary invariants only.

We have constructed a set of initial configurations consisting of a number
of twisted vortices where energy minimisation is known to lead to knotted
Hopfions in the $\sphere^2 \cartprod \torus^1$ case
\cite{Hietarinta:2003vn}. It is observed in this case, that in the
continuous deformation driven by energy minimization the conservation of
$s_1$ and $s_2$ does not allow for the unwinding of the knot and leads to
an intertwined vortex bunch. In the $\torus^3$ case neither the
conservation of $t_i$ nor energy minimisation requires that the knotted
structure remains intact and indeed the Hopfion unwinds. The conservation
of the homotopy invariants is confirmed by a visual inspection in all
cases.

The results have been compared with previously known deformation processes
of preimage splitting and reconnecting and found to follow the same
deformation ``rules'' as previously \cite{Hietarinta:2002xxx,
  Hietarinta:2003vn}. No new deformation processes were observed.

The results presented here have possible experimental relevance to the
observations of vortices in a rotating superfluid $^3$He in the $A$ phase
\cite{1995JETPL..61...49M, Ruutu:1995aa}. Another physically relevant
situation where maps fron $\torus^3 \to \sphere^2$ arise are insulating
magnetic materials in three dimensions, which can have topologically
nontrivial properties \cite{Moore:2008aa}.

\begin{acknowledgments}
  We gratefully acknowledge the generous computing resources of the Cray
  XT4 supercomputer at CSC~--~IT Center for Science Ltd. JJ has been
  supported by a research grant from the Academy of Finland (project
  123311).
\end{acknowledgments}

\bibliographystyle{apsrev}
\bibliography{bibliography}

\end{document}